\documentclass[aps,pre,twocolumn,showpacs,superscriptaddress,10pt,longbibliography]{revtex4-1}
\usepackage[T1]{fontenc}
\pdfoutput=1
\usepackage[dvipsnames,rgb,dvips]{xcolor}
\usepackage{float}
\usepackage{overpic}
\usepackage{bbm}
\usepackage{amsmath,amssymb,amsfonts}
\usepackage{hyperref}
\makeatletter
\def\amsbb{\use@mathgroup \M@U \symAMSb}
\makeatother
\usepackage{mathrsfs}
\makeatletter
\def\amsbb{\use@mathgroup \M@U \symAMSb}
\makeatother
\usepackage[bbgreekl]{mathbbol}
\newcommand{\T}{^{\sf T}}

\newcommand\ve[1]{\boldsymbol{#1}}
\newcommand{\ma}[1]{\ensuremath{\amsbb{#1}}}

\renewcommand{\tilde}{{}}
\usepackage{amsmath}
\usepackage{graphicx}
\usepackage{bbm}
\usepackage{bm}
\usepackage{subfig}
\usepackage{cleveref}
\usepackage{float}
\usepackage{caption}
\usepackage{booktabs}

\newcommand{\lp}{\ensuremath{\left (}}
\newcommand{\rp}{\ensuremath{\right )}}
\newcommand{\lsp}{\ensuremath{\left [}}
\newcommand{\rsp}{\ensuremath{\right ]}}

\newcommand{\e}{\ensuremath{\varepsilon}}

\begin{document}

\title{Angular dynamics of small crystals in viscous flow}

\author{J. Fries}
\affiliation{Department of Physics, University of Gothenburg, SE-41296 Gothenburg, Sweden}
\author{J. Einarsson}
\affiliation{Department of Physics, University of Gothenburg, SE-41296 Gothenburg, Sweden}
\author{B. Mehlig}
\affiliation{Department of Physics, University of Gothenburg, SE-41296 Gothenburg, Sweden}
\begin{abstract}
The angular dynamics of a very small ellipsoidal particle in a viscous flow decouples from its translational dynamics, and the particle angular velocity is given by Jeffery's theory. It is known that cuboid particles share these properties.  In the literature a special case is most frequently discussed, namely that of 
axisymmetric particles with a continuous rotation symmetry.
Here we compute the angular dynamics of crystals that possess a discrete 
rotation symmetry and certain mirror symmetries, but that do not have a continuous rotation symmetry. We give examples of such particles that nevertheless obey Jeffery's theory. But there are other examples where the angular dynamics is determined by a more general equation of motion. 
\end{abstract}
\maketitle

\section{Introduction}
Force and torque on a small particle in a viscous fluid depend linearly on its translational and angular slip velocities, and upon the local strain rate of the flow. In this \lq creeping-flow\rq{} limit, the particle moves so that the instantaneous force and torque vanish.
The constant coefficients in the linear law are given by the elements of the resistance tensors of the particle \cite{Happel:1983,Kim:2005}.

For an ellipsoidal particle, the elements of the resistance tensors can be 
deduced from the work of Jeffery \cite{Jeffery:1922} who 
derived an approximate 
equation of motion for such a particle: the translational motion of the particle does not affect its angular dynamics,
and the angular dynamics is determined by two shape parameters corresponding to the two aspect ratios that define the ellipsoid. 

Many studies of the angular motion of small non-spherical particles in flows concentrate on a special case of Jeffery's theory, spheroidal particles, that possess an axis of continuous rotation symmetry. In this case the angular dynamics is determined by a single shape parameter, the aspect ratio of the spheroid. Two examples are the angular dynamics of spheroidal particles in turbulence \cite{Par12,Gus14,Ni14,Che13,Byron2015,Zha15,Voth16}, and the rotation of a  spheroid           in a simple shear.  The second problem was solved by Jeffery \cite{Jeffery:1922} in the creeping-flow limit. He showed
that the tumbling dynamics has infinitely many, marginally stable periodic orbits (\lq Jeffery orbits\rq{}). This degeneracy means that small perturbations are important, such as rotational Brownian motion \cite{Hinch1972} and inertia \cite{subramanian2005,einarsson2014,einarsson2015b,candelier2015b,rosen2015d,Meibohm2016}. Bretherton \cite{Bretherton:1962} could show that particles with a continuous rotation symmetry have the same angular dynamics as spheroids,
in the creeping-flow limit.

What is known for other particle shapes? For any given particle shape the resistance tensors can be found by solving the Stokes problem determined by the particle shape, its orientation, and by the undisturbed flow 
as boundary conditions. An alternative is to use the symmetries of the problem to find conditions on the elements of the resistance tensors that allow to deduce the form of the equations of motion. In this way it can be shown that translational and angular dynamics decouple for \lq orthotropic\rq{} particles,
that is for particles with three mutually orthogonal mirror planes \cite{Happel:1983}. 
It is known that cuboids obey Jeffery's angular equation of motion with two shape parameters. This was shown by the experiments of Harris and coworkers \cite{Harris1979}. But in general the angular
dynamics of orthotropic particles is more complicated. 
Bretherton's theory \cite{Bretherton:1962} predicts that the angular dynamics is
described by three shape parameters.
\begin{figure}[t]
\begin{overpic}[width=0.7\columnwidth]{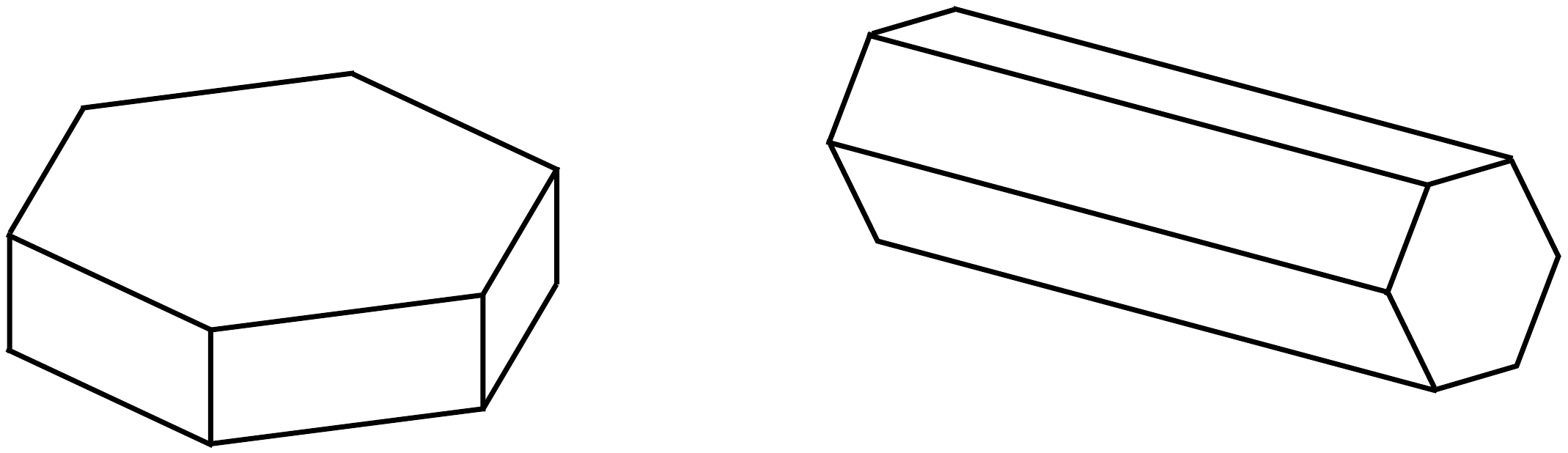}
\put(-6,25){\bf a}
\put(46,25){\bf b}
\end{overpic}
\caption{\label{fig:crystals} Examples of hexagonal ice crystals
(schematic)
with discrete rotation symmetry 
of order $6$. {\bf a}}~Hexagonal plate. {\bf b} Hexagonal column.  
\end{figure}

Most rigid particles that  we encounter in the natural world do not have the symmetries assumed above.
An example are crystals of hexagonal ice \cite{Nak54} that can have a $6$-fold discrete rotation symmetry (Fig.~\ref{fig:crystals}). Ice crystals play in an important part in rain initiation in cold cumulus clouds \cite{Mason1971,Pru78}. The tumbling dynamics of ice crystals in fully developed turbulence is therefore a question of current interest.
A second example is the angular dynamics of plankton in steady
and unsteady shear flows \cite{Gua12}. For microorganisms with inhomogeneous mass densities, fluid-velocity gradients 
and the gravitational torque compete to determine the angular motion \cite{Kes85}. Recent studies of the angular motion of such organisms in turbulence assume that they are spherical \cite{Dur13}
or spheroidal \cite{Zha14,Gus16}. But the  microorganisms often have very regular geometric shapes. An example
is the algae {\em Triceratium} that may assume the shapes of flat equilateral triangles, squares, or pentagons \cite{AlgaeBase}.

\begin{figure}[t]
\begin{overpic}[width=0.7\columnwidth]{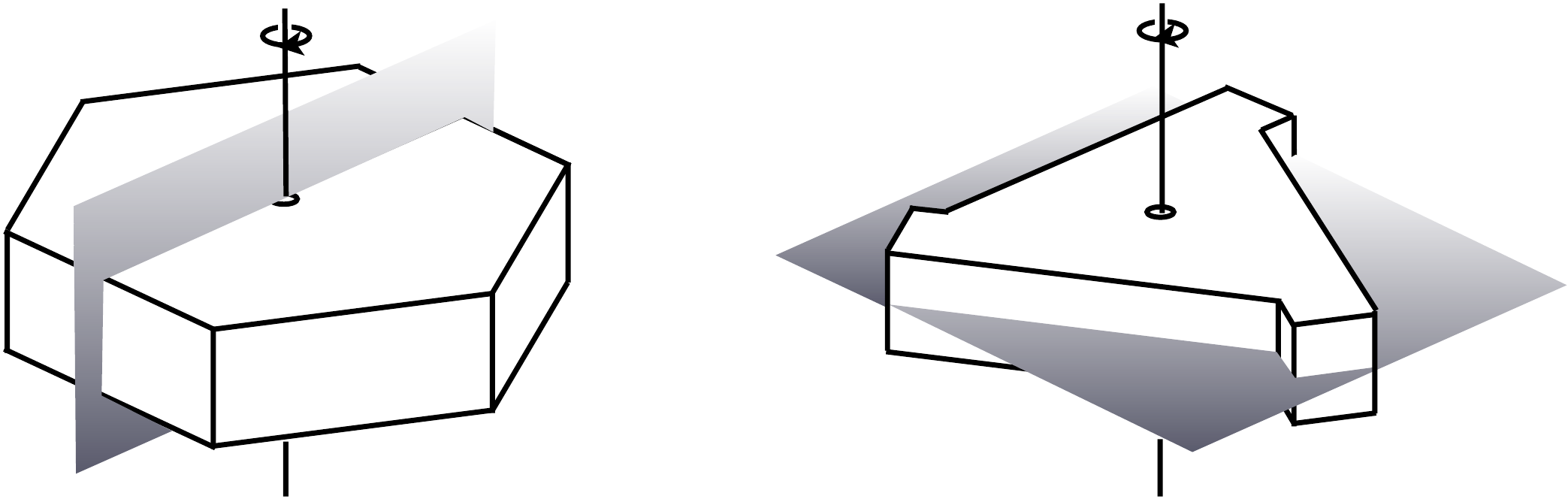}
\put(0,30){\bf a}
\put(55,30){\bf b}
\end{overpic}
\caption{\label{fig:sp} Mirror symmetries of particles with discrete rotation symmetry.
{\bf a} Mirror plane that contains the rotation-symmetry axis. 
{\bf b} Mirror plane orthogonal to the symmetry axis.}
\end{figure}

How can we parametrise the angular dynamics of such particles? To answer this question we analysed the angular dynamics of 
particles with a discrete $k$-fold rotation symmetry 
($k>2$), and in addition at least one mirror plane (Fig.~\ref{fig:sp}). Our analysis shows that the angular dynamics depends on whether the mirror plane contains the axis of rotation or not. 

If the mirror plane contains the axis of rotation (Fig. \ref{fig:sp}{\bf a}) then we conclude that the angular dynamics of the particle obeys Jeffery's equation for a spheroid. For the cuboid with at least one square face ($k=4$) this was known \cite{Bretherton:1962}, but for other values of $k$ our result generalises the symmetry considerations of Bretherton to particles with $k$-fold discrete rotation symmetry. 

If the mirror plane does not contain the axis 
of symmetry then this axis must be orthogonal to the mirror plane (Fig.~\ref{fig:sp}{\bf b}). The angular dynamics of such particles decouples from their translation dynamics. We derive the angular equation of motion, and show that the angular dynamics is parametrised by at most three dimensionless parameters.  We give an example of a particle for which the angular dynamics is characterised by two shape-dependent parameters. 
One parameter describes how the vector $\ve n_3$ aligned with the discrete rotation-symmetry axis tumbles. We find that it tumbles precisely like  a Jeffery spheroid.  The second parameter determines how the particle spins around this axis. We show that this motion is quite different from that predicted by Jeffery's theory for spheroids or ellipsoids.

\section{Background}
\label{sec:bg}
We consider a small particle in an incompressible viscous fluid
with undisturbed fluid velocity $\ve U^\infty(\ve x,t)$. 
The symmetric part of the matrix of fluid-velocity gradients $\partial U^\infty_i/\partial x_j$ is denoted by $\ma S^\infty(\ve x,t)$. This is the strain-rate matrix. 
The antisymmetric part of the velocity-gradient matrix is denoted by $\ma O^\infty(\ve x,t)$. This matrix 
characterises the flow rotation. It is related to half the the fluid vorticity,
 $\ve \Omega^\infty \equiv \tfrac{1}{2}\ve \nabla
\wedge \ve u$, by $\Omega^\infty_i = -\tfrac{1}{2}\e_{ijk}O_{jk}$. 
Here $\varepsilon_{ijk}$ is the Levi-Civit\`a{} tensor, and repeated indices are summed from $1$ to $3$. Below we also use the scalar
product $\ve a \cdot \ve b = a_i b_i$ between two vectors $\ve a$ and $\ve b$ as well as the Kronecker tensor $\delta_{ij}$.
We assume that the particle is so small that the unperturbed flow can be linearised around the particle position, and that
inertial effects are negligible. We also disregard the effect of molecular diffusion. In this limit the force $\ve F$ and torque  $\ve \tau$ acting upon a particle with translational velocity $\ve v$ and angular velocity $\ve \omega$ are given by \cite{Kim:2005}:
\begin{align}\label{tensors}
\begin{bmatrix}
\ve F \\
\ve \tau
\end{bmatrix}
= \mu 
\begin{bmatrix}
\ma A & {\ma B}^{\sf T} & \tilde{\ma G} \\
\ma B & \ma C & \tilde{\ma H} 
\end{bmatrix}
\begin{bmatrix}
\ve U^\infty_p - \ve v \\
\ve \Omega^\infty_p-\ve \omega \\
\ma S_p^\infty       
\end{bmatrix}.
\end{align}
Our notation differs slightly from that of Ref.~\cite{Kim:2005}.
In our Eq.~(\ref{tensors}), $\ve U^\infty_p$ stands for  the velocity of the undisturbed flow at the particle position, $\ve \Omega^\infty_p$ is half of the the undisturbed vorticity at the particle position, and $\ma S^\infty_p$ the strain-rate matrix at this position. The dynamic viscosity of the fluid is denoted by $\mu$.
Furthermore $\ma A$, $\ma B$ and $\ma C$ are rank-$2$ tensors, 
$\sf T$ denotes the transpose,
and $\tilde{\ma G}$ and $\tilde{\ma H}$ are rank-$3$ tensors. 
The product between a rank-$3$ and a rank-$2$ tensor is the double
contraction, e.g. $\tilde{\ma H}:\ma S_p^\infty$ with components
$(\tilde{\ma H}:\ma S_p^\infty)_i \equiv (\tilde{\ma  H})_{ijk} (\ma S_p^\infty)_{jk}$. 

\begin{figure}
\begin{overpic}[width=0.35\columnwidth]{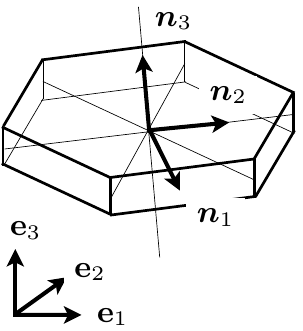}
\end{overpic}
\caption{\label{fig:coordinates}
Coordinate systems.  Particle-fixed coordinate system:  right-handed orthonormal basis
$\ve n_1$, $\ve n_2$, and $\ve n_3$. The vector $\ve n_3$ points along
the axis of discrete rotation symmetry.
The basis
vectors of the lab-fixed coordinate system
are denoted by  ${\bf e}_1$,  ${\bf e}_2$, and ${\bf e}_3$.}
\end{figure}

Eq.~(\ref{tensors}) is written in tensorial form, valid in this form independent of the choice of coordinate system. In this paper we use two different
coordinate systems  (Fig.~\ref{fig:coordinates}) to write down and analyse the elements of the resistance tensors: the lab-fixed basis ${\bf e}_j$ translates with the particle, but its orientation remains fixed in space. The particle-fixed basis $\ve n_\alpha$ also rotates with the particle.
It is important to consider the point with respect to which the torque and the resistance tensors are defined. The \lq centre of reaction\rq{} of the particle is uniquely defined as the point where $\ma B$ is symmetric \cite{Happel:1983}. 
For particles with certain rotation and mirror symmetries 
it can be shown \cite{Happel:1983} that this means $\ma B = 0$.  Eq.~(\ref{tensors}) then implies that translational and angular motion decouple.
The tensors $\ma A$ and $\ma C$ are symmetric regardless of whether they are defined with respect to the centre of reaction or not \cite{Happel:1983}. Also, $\ma A$ and $\ma C$ are positive definite \cite{Kim:2005}.

In the creeping-flow limit the angular equation of motion is derived as follows.
One expresses angular velocity and torque with respect to the centre of reaction. Then one
sets $\ve \tau=0$ in Eq.~(\ref{tensors}) to obtain an expression
for the angular velocity. Since $\ma C$ is a symmetric positive-definite matrix its inverse exists
and one finds \cite{Kim:2005}:
\begin{align}
\label{omega}
\ve\omega = \ve \Omega^\infty_p + \ma C^{-1} \tilde{\ma H} : \ma S_p^\infty\,.
\end{align}
Third, the angular equation of motion follows from
\begin{align}
\label{ndot}
\dot {\ve n}_\alpha = \ve \omega\wedge\ve n_\alpha
\end{align}
for $\alpha= 1,2,3$. The dot stands for  the time derivative of $\ve n_\alpha$ and the wedge denotes the cross product between $\ve \omega$ and $\ve n_\alpha$.
Jeffery solved the Stokes problem of an ellipsoidal particle in a simple shear,
calculated the particle angular velocity, Eq.~(\ref{omega}), and found
the angular equation of motion for an ellipsoidal particle. For the special case of a spheroid
this equation takes the form:
\begin{subequations}
\label{jeffern}
\begin{align}
\label{eq:n3}
\dot{\ve n}_3 &= \ma O_p^\infty \ve n_3 + \Lambda \lsp \ma S_p^\infty \ve n_3 
- \lp \ve n_3\cdot \ma S_p^\infty \ve n_3\rp \ve n_3\rsp\,,\\
\dot {\ve n}_1 &= \ma O_p^\infty \ve n_1 - \Lambda \lp \ve n_3\cdot \ma S_p^\infty 
\ve n_1\rp \ve n_3\,.
\end{align}
\end{subequations}
Here $\Lambda$ is the shape parameter, also
referred to as the \lq Bretherton constant\rq{}.
It is given by $\Lambda = (\lambda^2-1)/(\lambda^2+1)$, 
where $\lambda\equiv a/b$ is the aspect ratio of the spheroid
with half-axis lengths $a$ and $b$.
For $|\Lambda| < 1$ the solutions of Eq.~(\ref{eq:n3})
are the marginally stable periodic orbits known as Jeffery orbits. 

\section{Symmetry operations}\label{sec:transfrules}
The resistance tensors depend on the shape of the particle. 
Assume that the particle has certain reflection and rotation symmetries.
What does this imply
for the elements of $\ma B$, $\ma C$, and $\tilde{\ma H}$?

\begin{figure}[t]
\begin{overpic}[width=0.7\columnwidth]{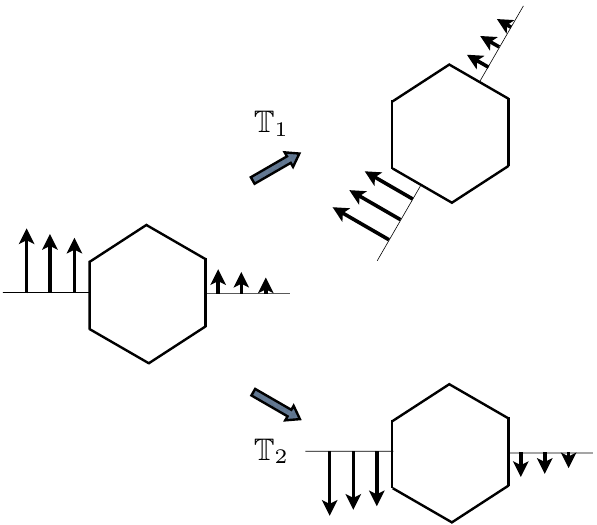}
\end{overpic}
\caption{\label{fig:transformation} Rotation of the flow with $\ma T_1$ results in the force $\ma T_1\ve F$ and the torque $\ma T_1\ve \tau$ since the orientation of the particle with respect to the flow remains unchanged. Reflection with $\ma T_2$ results in the force $\ma T_2\ve F$ and the torque $\det[\ma T_2]\,\ma T_2\ve \tau$. }
\end{figure}

The implications for $\ma B$ and $\ma C$ were described by Happel \& Brenner \cite{Happel:1983}. 
Consider the force and the torque upon a particle at
rest ($\ve v = 0$ and $\ve\omega = 0$).
Assume that the flow is transformed by an orthogonal transformation $\ma T$, that is a rotation (with
determinant $\det[\ma T] = 1$) or a reflection ($\det[\ma T] = -1$).  If the orientation of the particle surface in relation to the flow remains unchanged, then
the force on the particle is given by $\ma T\ve F$ (Fig.~\ref{fig:transformation}). The torque acquires a minus sign under reflections, therefore it must equal $\det[\ma T]\, \ma T\ve \tau$ after
transformation. Inserting this into Eq.~(\ref{tensors}) and using
the orthogonality of $\ma T$ one finds that the resistance
tensors $\ma B$ and $\ma C$ must satisfy
\begin{align}
\label{tensortransfBC}
\ma B &=  \det[\ma T]\,\ma T \ma B\ma T\T \quad 
\mbox{and}\quad \ma C =  \ma T \ma C\ma T\T\,.
\end{align}
What is the corresponding condition for $\tilde{\ma H}$? Bretherton answered this question for the special case of a particle that is symmetric under reflections w.r.t. two mutually orthogonal mirror planes (or equivalently $\pi/2$-rotations) \cite{Bretherton:1962}. 
In general $\tilde{\ma H}$ must obey an invariance relation analogous to (\ref{tensortransfBC}) under orthogonal transformations of the flow that leave the orientation of the particle surface with respect to the flow invariant. To derive this rule, assume that translational and angular slip vanish. In this case it follows from Eq.~(\ref{tensors})
that the torque is given by:
\begin{equation}
\label{eq:tau}
\ve \tau =\mu  \tilde{\ma H} : \ma S_p^\infty\,.
\end{equation}
Under an orthogonal transformation, $\ma S_p^\infty$ transforms
to  $\ma T \ma S_p^\infty\ma T\T$. 
If this transformation
leaves the orientation of the particle surface in relation to the flow invariant then the transformed flow must cause the torque $\det[\ma T]\ma T\ve \tau$,
\begin{equation}
\label{eq:tau2}
\det[\ma T]\ma T\ve \tau =\mu \tilde{\ma H}:\ma T \ma S_p^\infty \ma T\T\,.
\end{equation}
Now we eliminate $\ve \tau$ from Eqs.~(\ref{eq:tau})  and (\ref{eq:tau2}) and note
that the resulting equation must be valid for any rate-of-strain matrix
$\ma S_p^\infty$.
This yields the desired invariance relation. 
It is simpler to quote this relation in components. 
We write the components of tensors with respect to the lab-fixed basis 
${\bf e}_j$ with Roman indices,
and the components with respect to the particle-fixed basis $\ve n_\alpha$ with Greek indices.
In the particle-fixed basis, the invariance relation reads:
\begin{align}
\label{tensortransf}
\tilde{H}_{\alpha\beta\gamma} = \det[\ma T]\,T_{\alpha\alpha'} 
T_{\beta \beta'} T_{\gamma\gamma'} \tilde{H}_{\alpha'\beta'\gamma'}\,.
\end{align}
Eq.~(\ref{tensortransf}) constrains the possible values of the elements of $\tilde{\ma H}$.
In Eq.~(\ref{tensors}) the tensor
$\tilde{\ma H}$ occurs contracted with $\ma S_p^\infty$. Since the strain-rate matrix is symmetric 
we may assume that $\tilde{\ma H}$ is symmetric in the last two indices:
\begin{equation}
\label{eq:inc}
\tilde H_{\alpha\beta\gamma} = \tilde H_{\alpha\gamma\beta}\,.
\end{equation} 
 We use the relations (\ref{tensortransfBC}), (\ref{tensortransf}), and (\ref{eq:inc}) to deduce constraints for the elements of $\ma B, \ma C$ and $\tilde{\ma H}$, given certain symmetries of the particle shape. 
This allows us to derive how the angular equation of motion depends on the particle shape. In next Section we summarise the results of our analysis, details are given in the appendix. 

\section{Angular equations of motion}
\label{sec:aem}
For the particle shown in Fig.~\ref{fig:sp}{\bf a} there is a mirror plane that contains the axis of discrete rotation symmetry.
In this case the tensor $\ma B$ can be taken to vanish, so that translational and angular dynamics decouple.  Furthermore, $\ma C$ is diagonal in the particle-fixed basis with elements $C_{11} = C_{22}$ and $C_{33}$, and $\tilde{\ma H}$  has four non-zero elements $\tilde H_{123} = \tilde H_{132}
= -\tilde H_{213}= -\tilde H_{231}$,
fully determined by the  single parameter $\tilde H_{123}$ (see appendix).  Using Eq.~(\ref{omega}) we can deduce the angular velocity of the particle:
\begin{equation}
\label{eq:wj}
\ve \omega = \ve \Omega_p^\infty  -\Lambda \lp \ma S_p^\infty \ve n_3 \rp \wedge \ve n_3\,,
\end{equation}
with
$\Lambda = -2\tilde{H}_{123}/C_{11}$.
Jeffery's equation (\ref{jeffern})
follows directly from Eqs.~(\ref{eq:wj}) and (\ref{ndot}).
In summary, particles with a $k$-fold rotation symmetry ($k>2$) and a mirror plane orthogonal to this axis tumble and spin like a spheroid.

Now consider the particle shown in Fig.~\ref{fig:sp}{\bf b}. 
This particle has a discrete rotation symmetry and a mirror symmetry in a plane orthogonal to the rotation axis. 
Evaluating $\ma B$ with respect to the point where the rotation axis intersects
the mirror plane, one finds that $\ma B=0$. The tensor $\ma C$ is diagonal in the particle-fixed basis with elements
$C_{11}=C_{22}$ and $C_{33}$, the same as before. But now the tensor $\tilde{\ma H}$ can have up to $11$ non-zero elements in the
particle-fixed basis, parametrised by $\tilde H_{113}$,  $\tilde H_{123}$,
 $\tilde H_{311}$, and  $\tilde H_{333}$ (see appendix). 
In this case we deduce that the angular velocity of the particle,
\begin{equation}
\label{eq:omega2}
\ve \omega \!= \!\ve \Omega_p^\infty\!-\!\Lambda (\ma S_p^\infty \ve n_3) \wedge \ve n_3\!+\!\Gamma (\ve n_3 \cdot \ma S_p^\infty \ve n_3) \ve n_3 + \Psi \ma S_p^\infty \ve n_3\,,
\end{equation}
is para\-me\-tri\-sed in terms of three dimensionless parameters: $\Psi$, $\Gamma$, 
and the Bretherton constant $\Lambda$. They are given by
\begin{align}
\Lambda &= -2  \tilde{H}_{123}/C_{11}\,, \nonumber\\
\label{eq:params}
\Psi &= 2    \tilde{H}_{113}/C_{11}\,,\\
\Gamma & =   (\tilde{H}_{333} - \tilde{H}_{311})/C_{33} - \Psi\,.\nonumber
\end{align}
Using Eq.~(\ref{ndot}) we find the equations of motion for $\ve n_\alpha$:
\begin{align}
\dot{\ve n}_\alpha &= \ma O_p^\infty \ve n_\alpha + \Lambda \lsp (\ve n_3\cdot\ve n_\alpha)\,\ma S_p^\infty \ve n_3 \!-\! (\ve n_\alpha\cdot \ma S_p^\infty \ve n_3) \,\ve n_3\rsp\nonumber\\
&+ \Gamma [( \ve n_3 \cdot \ma S_p^\infty \ve n_3 ) \,\ve n_3] \wedge \ve n_\alpha 
 +\Psi \lp \ma S_p^\infty \ve n_3 \rp \wedge \ve n_\alpha\,.
\end{align}
There are  two additional terms compared with Jeffery's equation (\ref{jeffern}), parametrised
by $\Psi$ and $\Gamma$. 

We emphasise that the symmetry arguments used in this Section determine 
that the angular dynamics is given by at most three parameters. But the arguments
do not yield the values of $\Psi$ and $\Gamma$. 
There could be other symmetries, that we have not considered, that may constrain the values that $\Psi$ and $\Gamma$ can assume. Therefore it is important to find examples of particles that have, for instance,
$\Gamma\neq 0$. This is the topic of the next Section.

\begin{figure}[t]
\begin{overpic}[width=0.6\columnwidth]{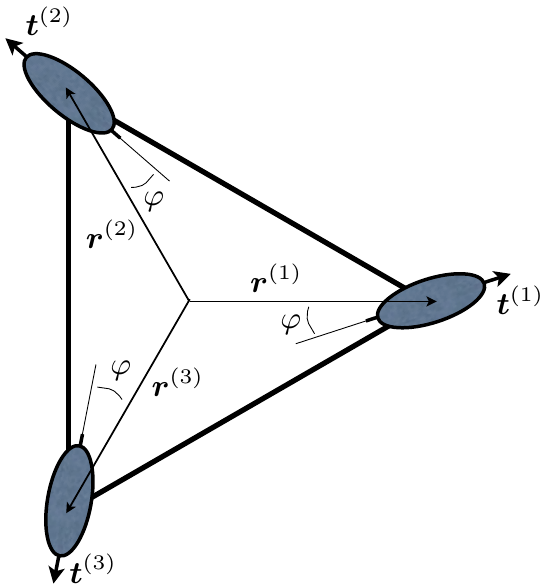}
\end{overpic}
\caption{\label{fig:triangle} Top view of a particle that has a $3$-fold rotation symmetry and a mirror plane orthogonal to the rotation axis. 
The particle consists of three spheroids connected by massless rigid rods that form an equilateral
triangle. The symmetry axes of the spheroids lie in the plane of the triangle and are rotated
(by the angle $\varphi$)
in such a way that the particle has $3$-fold discrete rotation symmetry but no mirror plane containing the rotation axis. 
The $\ve n_3$-vector points out of the image plane.}
\end{figure}

\section{A particle with $\Gamma\neq0$}
\label{sec:skewedtriangle}
Now we give an example for a particle that has $\Gamma\neq 0$, and $\Psi=0$.
The particle (Fig.~\ref{fig:triangle}) consists of three identical spheroids that are located at the corners of an equilateral triangle, linked by massless rigid rods. We assume that the distance between the spheroids is large in relation to their size, so that hydrodynamic interactions between them are negligible, as are the torques on individual spheroids.
Since we neglect hydrodynamic interactions we cannot construct a particle with nonzero $\Gamma$ or $\Psi$ by connecting spheres by massless rigid rods (by replacing each spheroid by a dumbbell for example). This follows from the fact that the forces on the spheres are additive. One possibility would be to construct a particle from slender rods \cite{Mar14}, but here we use spheroids.
Bretherton gave a formula for the torque on an assembly of spheroids, and used it to construct an example of a particle that is described by Jeffery's theory, but has $\Lambda > 1$ 
\cite{Bretherton:1962}. 

Here we consider the effect of breaking the mirror symmetry w.r.t.  a plane that contains the rotation axis. The symmetry is broken by rotating the spheroids by an angle $\varphi$. Fig.~\ref{fig:triangle} illustrates how this
angle $\varphi$ is defined. The spheroids are labeled by $s=1,2,3$, and
are located at $\ve r^{(s)}$ from the centre $\ve x_{\rm c}$ of the triangle. 
The torque with respect to the centre of the triangle is given by:
\begin{subequations}
\begin{eqnarray}
\ve \tau &=& \sum_{s=1}^3 \ve r^{(s)}\wedge \ve F^{(s)}\,,\\
\ve F^{(s)} &=& \mu\ma A^{(s)} [\ve U^\infty(\ve x_{\rm c}+\ve r^{(s)})-\ve v^{(s)}]\,.
\end{eqnarray}
\end{subequations}
Here $\ve v^{(s)}$ is the velocity of spheroid $s$, and $\ma A^{(s)}$ is its resistance to translational motion:
\begin{align}\label{rotresttens2}
\ma A^{(s)} &= \mathscr{A}_1\ma I + \mathscr{A}_2 \ve t^{(s)} \otimes \ve t^{(s)}\,,
\end{align}
where we have used the standard notation for the tensor product.
The parameters $\mathscr{A}_1$ and $\mathscr{A}_2$ are given by
the resistance functions of the spheroid, see 
Ref.~\cite{Kim:2005} or Table III in Ref.~\cite{einarsson2015b},
and $\ve t^{(s)}$ is the symmetry axis of spheroid~$s$.
We expand $\ve U^\infty$ around $\ve x_{\rm c}$.  
From Eq.~(\ref{tensors}) we can then read off the elements of the resistance tensors $\ma C$ and $\tilde{\ma H}$ that determine 
the dimensionless parameters $\Lambda$, $\Psi$, and $\Gamma$ of
the triangular particle:
\begin{eqnarray}
C_{11} &=& \tfrac{3}{2}a^2 \mathscr{A}_1\,,\quad C_{33} = 3a^2(\mathscr{A}_1+\mathscr{A}_2\sin^2\varphi)\\
\tilde{H}_{123}&=&\tfrac{3}{4}a^2\mathscr{A}_1\,,\,
\tilde{H}_{113}=0\,,\,\tilde{H}_{333}\!-\!\tilde{H}_{311} = -\tfrac{3}{4}a^2\mathscr{A}_2\sin2\varphi\,,
\nonumber
\end{eqnarray}
where $a\equiv|\ve r^{(s)}|$.
The elements are
expressed with respect to the centre of the triangle, in the particle-fixed basis.
From Eq.~(\ref{eq:params}) we find:
\begin{align}
\label{eq:Gam}
\Lambda = -1\,,\quad\Psi = 0\,,\quad\Gamma = -\tfrac{1}{4} 
\frac{\mathscr{A}_2  \sin 2 \varphi }{\mathscr{A}_1 + \mathscr{A}_2 \sin^2 \varphi}\,.
\end{align}
The fact that $\Lambda=-1$ is a consequence of the planar geometry of the particle,
and that the spheroids are much smaller than the side length of the triangle. A particle
with $|\Lambda |< 1$ and $\Gamma\neq 0$ can be constructed by stacking triangles on top of each other.

Eq.~(\ref{eq:Gam}) shows that $\Gamma = 0$ for $\varphi=0$ and $\pi/2$. This is consistent
with the conclusions drawn in Section \ref{sec:transfrules}, that a particle
with a mirror plane containing the rotation axis has $\Gamma=0$.
It follows from Eq.~(\ref{eq:Gam}) that $\Gamma$ is largest when $\varphi=\pi/4$. This is the value of $\varphi$ for which this mirror symmetry is broken most strongly, in the sense that the overlap between the particle and its reflected counterpart is minimal.

How does the $\Gamma$-term in Eq.~(\ref{eq:omega2}) affect the angular dynamics?
There are two contributions to the angular velocity, describing tumbling and  spinning of the particle: 
\begin{equation}
\omega^2 = \underbrace{\dot{n}_3^2}_{\rm tumbling} \!\!\!+ \,\,\,
\underbrace{(\ve \omega \cdot \ve n_3)}_{\rm spinning}\hspace*{-.3mm}\mbox{}^2\,.
\end{equation}
The $\Gamma$-term does not contribute to tumbling, $\dot{n}_3^2$. But
it does affect spinning, because
\begin{equation}
\label{eq:wn}
\ve\omega\cdot\ve n_3 = \ve \Omega_p^{\infty}\cdot \ve n_3 + \Gamma \,(\ve n_3\cdot\ma S_{{p}}^\infty\ve n_3)\,.
\end{equation}
Consider for example a simple shear with fluid-velocity gradients
$\partial U^\infty_i/\partial x_j = s \delta_{i1}\delta_{j2}$.
Eq.~(\ref{eq:wn}) shows that the effect of the $\Gamma$-term is 
largest when the $\ve n_3$-vector tumbles in the flow-shear plane, 
where $\ve \Omega_p^{\infty}\cdot \ve n_3=0$. The vorticity term
is largest when $\ve n_3$ aligns with the vorticity axis. In this case the $\Gamma$-term does not contribute to the spin.

\begin{figure}[t]
\begin{overpic}[width=0.7\columnwidth]{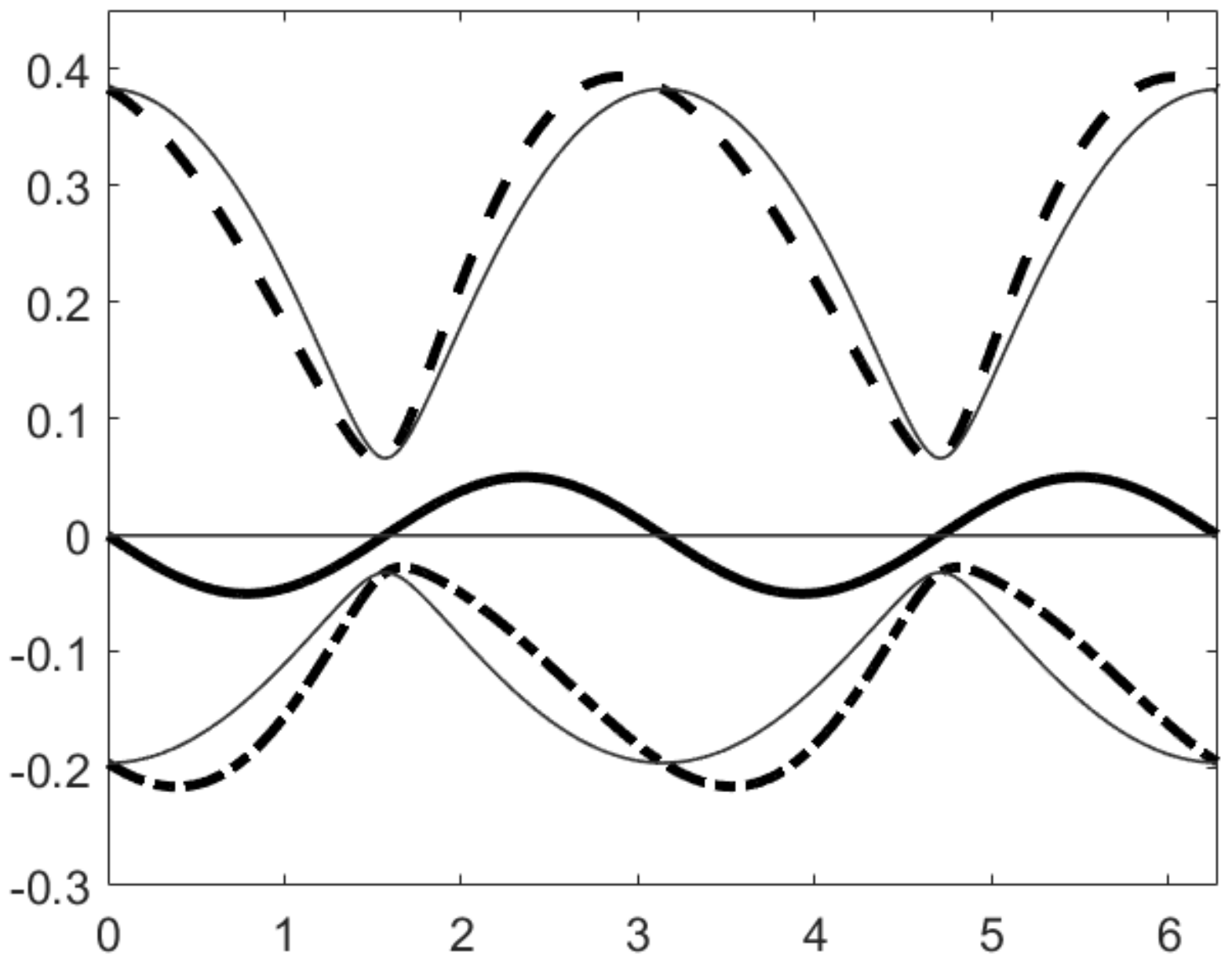}
\put(-4,40){\rotatebox{90}{$\ve \omega\cdot\ve n_3/s$}}
\put(55,-2){$-\phi$}
\end{overpic}
\caption{\label{fig:gamma} 
Spinning of a particle with $\Lambda=-0.95$, $\Gamma=0.1$,
and $\Psi=0$ in three different Jeffery orbits. The orbits
are parametrised by the precession angle $\phi$ and the polar angle
$\theta_0$ at $\phi=0$ (see text): $\theta_0= \pi/2$ (thick solid line),
$\theta_0 = 9\pi/16$ (thick dash-dotted line), and $\theta_0 = 3\pi/8$ (thick dashed line). Thin lines are for $\Gamma=0$.}
\end{figure}

These conclusions are illustrated in Fig.~\ref{fig:gamma}.  It shows how the spinning rate $\ve \omega \cdot \ve n_3$ varies for different Jeffery orbits for a particle with $\Lambda = -0.95$.  The orbits are parametrised by the \lq precession\rq{} angle $\phi$, defined by $\ve n_3 = [\cos\phi\sin\theta,\sin\phi\sin\theta,\cos\theta]^{\sf T}$, and $\theta$ is the polar angle, the angle between $\ve n_3$ and the ${\bf e}_3$-axis, the negative vorticity axis. Shown are three different orbits.
Thin lines show the results for $\Gamma=0$, and thick lines correspond to $\Gamma=0.1$. We see that the particle with $\Gamma=0.1$ spins quite differently from the particle with $\Gamma=0$, in agreement with our analysis of Eq.~(\ref{eq:wn}) summarised above.

\section{Conclusions}
\label{sec:conc}
In this paper we have analysed the angular dynamics of crystals with discrete rotation symmetries and certain reflection symmetries in the creeping-flow limit. We have used symmetry arguments to show that the particles shown in Fig.~\ref{fig:crystals} spin and tumble precisely like spheroids. In other words their angular dynamics is determined by Jeffery's theory \cite{Jeffery:1922}. But the particle shown in Fig.~\ref{fig:sp}{\bf b} spins differently, due to an additional term in the equation for the angular velocity of the particle. It appears because the particle does not have a reflection symmetry that contains the axis of discrete rotation symmetry. 

We remark that the particles shown in Figs.~\ref{fig:crystals} and \ref{fig:sp} are not chiral. Other authors have studied
the translational and angular dynamics of chiral particles in steady and turbulent flows, see for instance \cite{Mei12,Kra16,Gus16b}.

Our results raise interesting questions for further studies. How small particles spin and tumble in turbulence
is determined by the statistics of their alignment with vorticity, and this depends on the particle shape.
For spheroidal particles this question has been analysed in detail  \cite{Pum11,Par12,Gus14,Ni14,Che13}.
But particles with $\Gamma\neq 0$ exhibit different spinning statistics, and it would be of interest to perform direct numerical simulations of the angular
dynamics of such particles in turbulence. It was found in Ref.~\cite{Byron2015} that the mean-squared angular velocity of spheroids in turbulence
is essentially independent of the aspect ratio. Is this also true for particles like the one shown in Fig.~{\ref{fig:sp}{\bf b}}?  Such particles can be printed using a 3D printer, and their spinning can be analysed experimentally using the techniques described in Refs.~\cite{Byron2015,Ni15}. 

While the tumbling of spheroids is regular, the tumbling of ellipsoids can be chaotic. It turns out that the angular dynamics is very sensitive to a weak breaking of axisymmetry  \cite{Hinch1979,Yarin1997,einarsson2016a}, and that the effect of symmetry breaking on the angular dynamics is in many ways similar to a weak perturbation of an integrable Hamiltonian dynamical system \cite{einarsson2016a,Strogatz}, despite the fact that the angular dynamics is dissipative. 
 It would be of interest to analyse the effect of breaking the $k$-fold rotation symmetry when $\Gamma\neq 0$. The details of the transition to chaos depend on how tumbling and spinning couple, and thus on the spinning dynamics. The latter is affected by whether $\Gamma$ is zero or not. 

The triangular shapes of the diatom {\em Triceratium} mentioned in the Introduction are not perfectly symmetric. In this connection weak breaking of the discrete rotation symmetry is of interest. 
For the angular dynamics of ice crystals in turbulent clouds, inertial effects may be important. It is straightforward to take into account
particle inertia. Fluid-inertia effects are more difficult to treat.

\acknowledgements{We thank E. Variano for discussions and for directing us to Ref.~\cite{AlgaeBase}. We acknowledge  support by Vetenskapsr\aa{}det [grant number 2013-3992], Formas [grant number 2014-585],  by the grant \lq Bottlenecks for particle growth in turbulent aerosols\rq{} from the Knut and Alice Wallenberg Foundation, Dnr. KAW 2014.0048, by the Carl Trygger Foundation for Scientific Research, and by the MPNS COST Action MP1305 \lq Flowing matter\rq{}.}

\appendix
\section{Derivation of Eqs.~(\ref{eq:wj}), (\ref{eq:omega2}), and (\ref{eq:params})}
For the particle shown in Fig.~\ref{fig:sp}{\bf a} we consider
the following symmetry operations. Symmetry operation $1$ is a 
counterclockwise rotation around $\ve n_3$ by the angle $\alpha=2\pi/k$
for $k=6$ (Fig.~\ref{fig:transformation}, top). In the particle-fixed basis this transformation reads:
\begin{align}
\label{eq:s1}
\ma T_1 =
\begin{bmatrix}
\cos \alpha & -\sin \alpha & 0 \\
\sin \alpha & \cos \alpha & 0 \\
0 & 0 & 1
\end{bmatrix}\,.
\end{align}
Symmetry operation $2$ is a reflection in a plane that contains $\ve n_3$
(Fig.~\ref{fig:transformation}, bottom), given by the transformation matrix
\begin{align}
\label{eq:s2}
\ma T_2 =
\begin{bmatrix}
-1 & 0 & 0 \\
0 & 1 & 0 \\
0 & 0 & 1
\end{bmatrix}\,
\end{align}
in the particle-fixed basis.
The particle shown in Fig.~\ref{fig:sp}{\bf b} does not possess symmetry $2$, 
but it is invariant under
reflection in a plane that has $\ve n_3$ as a normal vector
(symmetry operation $3$). It is given by the transformation matrix
\begin{align}
\label{eq:s3}
\ma T_3 =
\begin{bmatrix}
1 & 0 & 0 \\
0 & 1 & 0 \\
0 & 0 & -1
\end{bmatrix}\,
\end{align}
in the particle-fixed basis.
In the following we examine the forms of $\ma B$, $\ma C$, and $\tilde{\ma H}$ for particles possessing 
the discrete symmetry $1$ for an angle $\alpha$ that is not a multiple of $\pi$, and either
symmetry $2$ or $3$.  

Consider first the particles shown in Fig.~\ref{fig:crystals}. They possess the symmetries 1 and 2. 
Using Eqs.~(\ref{tensortransfBC}), (\ref{tensortransf}), (\ref{eq:inc}), (\ref{eq:s1}), and (\ref{eq:s2}) 
one finds that the resistance tensors must be of the form:
\begin{align}
\label{eq:bc1}
& \ma B = \begin{bmatrix}
0 & B_{12} & 0\\
-B_{12} &  0 & 0\\
0 & 0 & 0
\end{bmatrix}
 , \,\,
\ma C = \begin{bmatrix}
C_{11} & 0 & 0 \\
0 & C_{11} & 0 \\
0 & 0 & C_{33}
\end{bmatrix},
\end{align}
and
\begin{align}
& \tilde{\ma H}_{1,:,:} = \begin{bmatrix}
0 & 0 & 0 \\
0 &  0 & \tilde{H}_{123} \\
0 & \tilde{H}_{123} & 0
\end{bmatrix},
& \tilde{\ma H}_{2,:,:} = \begin{bmatrix}
0 & 0 & -\tilde{H}_{123} \\
0 &  0 & 0 \\
-\tilde{H}_{123} & 0 & 0
\end{bmatrix},\,\nonumber\\
& \tilde{\ma H}_{3,:,:} = \begin{bmatrix}
0 & 0 & 0 \\
0 & 0 & 0 \\
0 & 0& 0
\end{bmatrix}\,,
\label{eq:h1}
\end{align}
in the particle-fixed basis.
We note that $\ma B$ comes out antisymmetric. However, there exists a point in the particle frame of reference at which $\ma B$ is symmetric, the centre of reaction \cite{Happel:1983}. This means that $\ma B = 0$ with resepct to  this point. We also see that the tensor $\ma C$ has two independent elements, and $\tilde{\ma H}$ just one. Below         we demonstrate that the corresponding angular dynamics is given by Eq.~(\ref{jeffern}), Jeffery's equation for a spheroid.

Now consider the particle shown in Fig.~\ref{fig:sp}{\bf b}. 
Invoking Eqs.~(\ref{tensortransfBC}), (\ref{tensortransf}), (\ref{eq:inc}), (\ref{eq:s1}), and (\ref{eq:s3})
we find
\begin{align}
& \ma B = \begin{bmatrix}
0 & 0 & 0\\
0 &  0 & 0\\
0 & 0 & 0
\end{bmatrix},\,
\ma C = \begin{bmatrix}
C_{11} & 0 & 0 \\
0 & C_{11} & 0 \\
0 & 0 & C_{33}
\end{bmatrix}\,,
\end{align}
and
\begin{align}
\nonumber
& \tilde{\ma H}_{1,:,:} = \begin{bmatrix}
0 & 0 & \tilde{H}_{113} \\
0 &  0 & \tilde{H}_{123} \\
\tilde{H}_{113} & \tilde{H}_{123} & 0
\end{bmatrix},\\\label{rotnh}
& \tilde{\ma H}_{2,:,:} = \begin{bmatrix}
0 & 0 & -\tilde{H}_{123} \\
0 &  0 & \tilde{H}_{113} \\
-\tilde{H}_{123} & \tilde{H}_{113} & 0
\end{bmatrix},\\
& \tilde{\ma H}_{3,:,:} = \begin{bmatrix}
\tilde{H}_{311} & 0 & 0 \\
0 & \tilde{H}_{311} & 0 \\
0 & 0& \tilde{H}_{333}
\end{bmatrix}
\nonumber
\end{align}
in the particle-fixed basis. The symmetry $3$ does not constrain the elements of $\tilde{\ma H}$ further than symmetry $1$, but symmetry $3$ 
allows us to conclude that $\ma B = 0$. 

Now we use the forms of the tensors obtained above to
derive the angular equation of motion. This involves two steps.
First we write the tensors in basis-independent form and then determine their components in the lab-fixed coordinate system. Second we use 
Eqs.~(\ref{omega}) and (\ref{ndot}).

\subsection{Symmetries 1 and 2}
\label{sec:s12}
In Eqs.~(\ref{eq:bc1}) and (\ref{eq:h1}) 
the tensors $\ma C$ and $\tilde{\ma H}$ 
are written in the particle-fixed basis. 
The basis-independent form
of $\ma C$ can be read off from (\ref{eq:bc1}):
\begin{align}\label{rotrestens}
\ma C &= C_{\alpha\beta} \,\ve n_\alpha\otimes \ve n_\beta \nonumber \\
&= C_{11} (\ve n_1\otimes \ve n_1 + \ve n_2\otimes \ve n_2)
+ C_{33} \ve n_3\otimes \ve n_3\\
&= C_{11} (\ma I-\ve n_3\ve n_3\T)
+ C_{33} \ve n_3 \ve n_3\T\nonumber\,.
\end{align}
To write the tensor  $\tilde{\ma H}$ in terms of $\ve n_\alpha$
we use:
\begin{align}
\label{eq:tensorH}
\tilde{\ma H} &= {\tilde H}_{\alpha\beta\gamma}\,\det[\ve n_1,\ve n_2,\ve n_3]\,
\ve n_\alpha \otimes \ve n_\beta \otimes \ve n_\gamma\,.
\end{align}
The determinant ensures that $\tilde{\ma H}$ changes sign upon reflection, it transforms as
$H'_{\alpha\beta\gamma} = \det[\ma T]$ $T_{\alpha\alpha'} T_{\beta\beta'}T_{\gamma\gamma'}$ $H_{\alpha'\beta'\gamma'} = H_{\alpha\beta\gamma}(\ma T \ve n_\delta)$.
This is consistent with the invariance relation (\ref{tensortransf}).
Using Eqs.~(\ref{eq:h1}) and (\ref{eq:tensorH}) we evaluate 
the components of $\tilde{\ma H}$ in the lab-fixed basis:
\begin{align}
\label{htrans1}
\tilde{H}_{ijk} = \tilde{H}_{123} (\ve z\cdot\ve n_3)\,  \tilde{H}_{123}\,
(\varepsilon_{rij}  n_{3k} + \varepsilon_{rik}  n_{3j}) z_r\,,
\end{align}
with $\ve z \equiv \ve n_1 \wedge \ve n_2$ so that 
$\det[\ve n_1,\ve n_2,\ve n_3]=\ve z\cdot\ve n_3$.
Eq.~(\ref{htrans1}) is equivalent to the form of
$\tilde{\ma H}$ for a spheroid \cite{Kim:2005}.  We have thus shown that particles with a $k$-fold
rotation symmetry ($k>2$) and a mirror plane orthogonal to this axis tumble and spin like a spheroid.  
To obtain the corresponding angular velocity in explicit form we use 
Eqs.~(\ref{omega}),  (\ref{rotrestens}) and (\ref{htrans1}). This gives:
\begin{equation}
\ve \omega = \ve \Omega_p^\infty  -\Lambda \lp \ma S_p^\infty \ve n_3 \rp \wedge \ve z
\end{equation}
with
\begin{equation}
\Lambda = -2(\ve z\cdot\ve n_3)\tilde{H}_{123}/C_{11}\,.
\end{equation} 
To make the symmetry properties of these equations explicit we have
kept the notation $\ve z= \ve n_1\wedge \ve n_2$. In the main text
we have replaced $\ve z$ by $\ve n_3$. This simplifies the equation, 
Eq.~(\ref{eq:wj}), yields the correct angular dynamics, but  breaks the transformation properties under reflection.

\subsection{Symmetries 1 and 3}
We find the components
of $\tilde{\ma H}$ in the lab-fixed basis using
Eqs.~(\ref{rotnh}) and (\ref{eq:tensorH}):
\begin{align}
\begin{split}
\label{eq:hijk}
\tilde{H}_{ijk} &= (\ve z\cdot\ve n_3) [(\tilde{H}_{333}-2\tilde{H}_{113}-\tilde{H}_{311}) n_{3i} n_{3j} n_{3k}\\
&+\tilde{H}_{113}\delta_{ij}n_{3k}+\tilde{H}_{311}\delta_{jk}n_{3i} +\tilde{H}_{113}\delta_{ki}n_{3j}\\
&+\tilde{H}_{123} \lp \e_{ijr}n_{3k}+\e_{ikr} n_{3j} \rp {z_r}]\,.
\end{split}
\end{align}
The angular velocity follows using incompressibility of the flow, $S_{ii}=0$,
and Eqs.~(\ref{omega}), (\ref{rotrestens}), and (\ref{eq:hijk}):
\begin{equation}
\ve \omega \!= \!\ve \Omega_p^\infty\!-\!\Lambda (\ma S_p^\infty \ve n_3) \wedge \ve z\!+\!\Gamma (\ve n_3 \cdot \ma S_p^\infty \ve n_3) \ve n_3 + \Psi \ma S_p^\infty \ve n_3\,.
\end{equation}
\mbox{}\\[-3mm]
The angular velocity is para\-me\-tri\-sed in terms of three dimensionless parameters: $\Psi$, $\Gamma$, 
and the Bretherton constant $\Lambda$. They are given by
\begin{align}
\Lambda &= -2  (\ve z\cdot\ve n_3) \tilde{H}_{123}/C_{11}\,, \nonumber\\
\Psi &= 2  (\ve z\cdot\ve n_3)  \tilde{H}_{113}/C_{11}\,,\\
\Gamma & =  (\ve z\cdot\ve n_3)  (\tilde{H}_{333} - \tilde{H}_{311})/C_{33} - \Psi\,.\nonumber
\end{align}
Upon substituting $\ve n_3$ for $\ve z$ we obtain Eqs.~(\ref{eq:omega2}) and (\ref{eq:params}).
\vfill

\end{document}